# Assemblathon 2: evaluating *de novo* methods of genome assembly in three vertebrate species

## Author information

Keith R. Bradnam[1*+]
[*] Corresponding author
[+] These authors contributed equally to this work
Email: krbradnam@ucdavis.edu

Joseph N. Fass[1+]
[+] These authors contributed equally to this work
Email: jnfass@ucdavis.edu

Anton Alexandrov[36]
Email: alantbox@gmail.com

Paul Baranay[2]
Email: pbaranay@cshl.edu

Michael Bechner[39]
Email: mbechner@lmcg.wisc.edu

Inanç Birol[33]
Email: ibirol@bcgsc.ca

Sébastien Boisvert[10,11]
Email: sebastien.boisvert.3@ulaval.ca

Jarrod A. Chapman[20]
Email: jchapman@lbl.gov

Guillaume Chapuis[7,9]
Email: guillaume.chapuis@irisa.fr

Rayan Chikhi[7,9]
Email: chikhi@irisa.fr

Hamidreza Chitsaz[6]
Email: chitsaz@wayne.edu

Wen-Chi Chou[14,16]
Email: wcc@uga.edu

Jacques Corbeil[10,13]
Email: jacques.corbeil@crchul.ulaval.ca

Cristian Del Fabbro[17]
Email: delfabbro@appliedgenomics.org

T. Roderick Docking[33]
Email: rdocking@bcgsc.ca

Richard Durbin[34]
Email: rd@sanger.ac.uk

Dent Earl[40]
Email: dearl@soe.ucsc.edu

Scott Emrich[3]
Email: semrich@nd.edu

Pavel Fedotov[36]
Email: fedotovp@gmail.com

Nuno A. Fonseca[30, 35]
Email: nuno.fonseca@gmail.com

Ganeshkumar Ganapathy[38]
Email: gg28@duke.edu

Richard A. Gibbs[32]
Email: agibbs@bcm.edu

Sante Gnerre[22]
Email: sante@broadinstitute.org

Élénie Godzaridis[11]
Email: elenie.godzaridis.1@ulaval.ca

Steve Goldstein[39]
Email: steveg@lmcg.wisc.edu

Matthias Haimel[30]

Email: mh719@cam.ac.uk

Giles Hall[22]
Email: giles@polymerase.org

David Haussler[40]
Email: haussler@soe.ucsc.edu

Joseph B. Hiatt[41]
Email: jbhiatt@uw.edu

Isaac Y. Ho[20]
Email: isaacyho@gmail.com

Jason Howard[38]
Email: howard@neuro.duke.edu

Martin Hunt[34]
Email: mh12@sanger.ac.uk

Shaun D. Jackman[33]
Email: sjackman@bcgsc.ca

David B Jaffe[22]
Email: jaffe@broadinstitute.org

Erich Jarvis[38]
Email: jarvis@neuro.duke.edu

Huaiyang Jiang[32]
Email: jiangh3@mwri.magee.edu

Sergey Kazakov[36]
Email: kazakov_sergey_v@mail.ru

Paul J. Kersey[30]
Email: pkersey@ebi.ac.uk

Jacob O. Kitzman[41]
Email: kitz@uw.edu

James R. Knight[37]
Email: james.knight@roche.com

Sergey Koren[24, 25]
Email: sergekoren@gmail.com

Tak-Wah Lam[29]
Email: twlam@cs.hku.hk

Dominique Lavenier[7,8,9]
Email: dominique.lavenier@irisa.fr

François Laviolette[12]
Email: francois.laviolette@ift.ulaval.ca

Yingrui Li[28, 29]
Email: liyr@genomics.org.cn

Zhenyu Li[28]
Email: lizhenyu@genomics.org.cn

Binghang Liu[28]
Email: liubinghang@genomics.org.cn

Yue Liu[32]
Email: yl131317@bcm.edu

Ruibang Luo[28,29]
Email: luoruibang@genomics.org.cn

Iain MacCallum[22]
Email: iainm@broadinstitute.org

Matthew D MacManes[5]
Email: macmanes@gmail.com

Nicolas Maillet[8,9]
Email: nicolas.maillet@inria.fr

Sergey Melnikov[36]
Email: svembox@gmail.com

Delphine Naquin[8,9]
Email: delphine.naquin@cgm.cnrs-gif.fr

Zemin Ning[34]
Email: zn1@sanger.ac.uk

Thomas D. Otto[34]
Email: tdo@sanger.ac.uk

Benedict Paten[40]
Email: benedict@soe.ucsc.edu

Octávio S. Paulo[31]
Email: octavio.paulo@fc.ul.pt

Adam M. Phillippy[24, 25]
Email: aphillippy@gmail.com

Francisco Pina-Martins[31]
Email: f.pinamartins@gmail.com

Michael Place[39]
Email: mplace@wisc.edu

Dariusz Przybylski[22]
Email: dariusz@broadinstitute.org

Xiang Qin[32]
Email: xqin@bcm.edu

Carson Qu[32]
Email: jqu@bcm.edu

Filipe J Ribeiro[22]
Email: fjribeiro@nygenome.org

Stephen Richards[32]
Email: stephenr@bcm.edu

Daniel S. Rokhsar[20, 21]
Email: dsrokhsar@gmail.com

J. Graham Ruby[26, 27]
Email: grahamruby@yahoo.com

Simone Scalabrin[17]
Email: scalabrin@appliedgenomics.org

Michael C. Schatz[4]

Email: mschatz@cshl.edu

David C. Schwartz[39]
Email: dcschwartz@facstaff.wisc.edu

Alexey Sergushichev[36]
Email: alsergbox@gmail.com

Ted Sharpe[22]
Email: tsharpe@broadinstitute.org

Timothy I. Shaw[14,15]
Email: gatechatl@gmail.com

Jay Shendure[41]
Email: shendure@uw.edu

Yujian Shi[28]
Email: shiyujian@genomics.org.cn

Jared T. Simpson[34]
Email: js18@sanger.ac.uk

Henry Song[32]
Email: XSong3@mdanderson.org

Fedor Tsarev[36]
Email: fedor.tsarev@gmail.com

Francesco Vezzi[19]
Email: francesco.vezzi@scilifelab.se

Riccardo Vicedomini[17,18]
Email: rvicedomini@appliedgenomics.org

Bruno M. Vieira[31]
Email: bmpvieira@gmail.com

Jun Wang[28]
Email: wangj@genomics.org.cn

Kim C. Worley[32]
Email: kworley@bcm.edu

Shuangye Yin[22]
Email: shuangye@broadinstitute.org

Siu-Ming Yiu[29]
Email: smyiu@cs.hku.hk

Jianying Yuan[28]
Email: yuanjianying@genomics.org.cn

Guojie Zhang[28]
Email: zhanggj@genomics.org.cn

Hao Zhang[28]
Email: zhanghao2@genomics.org.cn

Shiguo Zhou[39]
Email: szhou@wisc.edu

Ian F. Korf[1*]
Email: ifkorf@ucdavis.edu
[*] Corresponding author

**Affiliations**

1 Genome Center, UC Davis, CA, 95616
2 Computational Biology and Bioinformatics, Yale University, New Haven, CT 06511
3 Department of Computer Science and Engineering, University of Notre Dame, South Bend, IN, 46556
4 Simons Center for Quantitative Biology, Cold Spring Harbor Laboratory, Cold Spring Harbor, NY 11724
5 University of California, Berkeley California Institute for Quantitative Biosciences Berkeley, CA. USA 94720
6 Department of Computer Science, Wayne State University
7 Computer Science department, ENS Cachan/IRISA, 35042 Rennes, France
8 INRIA, Rennes Bretagne Atlantique, 35042 Rennes, France
9 CNRS/Symbiose, IRISA, 35042 Rennes, France
10 Infectious Diseases Research Center, CHUQ Research Center, Québec, QC, Canada
11 Faculty of Medicine, Laval University, Québec, QC, Canada
12 Department of Computer Science and Software Engineering, Faculty of Science and Engineering, Laval University, Québec, QC, Canada
13 Department of Molecular Medicine, Faculty of Medicine, Laval University, Québec, QC, Canada
14 Institute of Bioinformatics, University of Georgia, Athens, GA USA
15 Department of Epidemiology and Biostatistics, College of Public Health, University of Georgia, Athens, GA USA
16 Institute of Aging Research, Hebrew SeniorLife, Boston, MA USA
17 IGA, Institute of Applied Genomics, 33100 Udine, Italy
18 Department of Mathematics and Computer Science, University of Udine, 33100 Udine, Italy
19 Science for Life Laboratory, KTH Royal Institute of Technology, 17121 Solna, Sweden
20 DOE Joint Genome Institute, Walnut Creek, CA
21 UC Berkeley, Dept, of Molecular and Cell Biology, Berkeley, CA
22 Broad Institute, Cambridge, Massachusetts 02142, USA
23 New York Genome Center, New York, NY 10022, USA
24 National Biodefense Analysis and Countermeasures Center, Frederick, MD
25 Center for Bioinformatics and Computational Biology, University of Maryland, College Park, MD
26 Department of Biochemistry and Biophysics, University of California, San Francisco, California 94143, USA
27 Howard Hughes Medical Institute, Bethesda, Maryland 20814, USA
28 BGI-Shenzhen, China.
29 HKU-BGI Bioinformatics Algorithms and Core Technology Research Laboratory.
30 EMBL-European Bioinformatics Institute, Wellcome Trust Genome Campus, Hinxton, Cambridge CB10 1SD, UK
31 Computational Biology & Population Genomics Group, Centre for Environmental Biology, Department of Animal Biology, Faculty of Sciences of the University of Lisbon, Campo Grande, P-1749-016 Lisbon, Portugal 32 Human Genome Sequencing Center and Department of Molecular and Human Genetics, Baylor College of Medicine, Houston, TX 77030, USA
33 Genome Sciences Centre, British Columbia Cancer Agency, Vancouver, British Columbia, Canada V5Z 4E6
34 The Wellcome Trust Sanger Institute, Wellcome Trust Genome Campus, Hinxton, Cambridge, CB10 1SA, UK
35 CRACS - INESC TEC, Porto, Portugal
36 National Research University of Information Technology, Mechanics and Optics (University ITMO), St. Petersburg, Russia
37 454 Life Sciences. 15 Commercial Street, Branford, CT 06405, USA
38 Duke University Medical Center, Durham, NC 27710
39 Laboratory for Molecular and Computational Genomics, Departments of Chemistry and Genetics, UW-Biotechnology Center, 425 Henry Mall, Madison, Wisconsin 53706, USA
40 Howard Hughes Medical Institute, Center for Biomolecular Science & Engineering, University of California, Santa Cruz, CA 95064, USA
41 Department of Genome Sciences, School of Medicine, University of Washington, Seattle, WA 98195, USA

# Abstract

**Background**

The process of generating raw genome sequence data continues to become cheaper, faster, and more accurate. However, assembly of such data into high-quality, finished genome sequences remains challenging. Many genome assembly tools are available, but they differ greatly in terms of their performance (speed, scalability, hardware requirements, acceptance of newer read technologies) and in their final output (composition of assembled sequence). More importantly, it remains largely unclear how to best assess the quality of assembled genome sequences. The Assemblathon competitions are intended to assess current state-of-the-art methods in genome assembly.

**Results**

In Assemblathon 2, we provided a variety of sequence data to be assembled for three vertebrate species (a bird, a fish, and snake). This resulted in a total of 43 submitted assemblies from 21 participating teams. We evaluated these assemblies using a combination of optical map data, Fosmid sequences, and several statistical methods. From over 100 different metrics, we chose ten key measures by which to assess the overall quality of the assemblies.

**Conclusions**

Many current genome assemblers produced useful assemblies, containing a significant representation of their genes and overall genome structure. However, the high degree of variability between the entries suggests that there is still much room for improvement in the field of genome assembly and that approaches which work well in assembling the genome of one species may not necessarily work well for another.

# Background

Continued advances in next-generation sequencing (NGS) technologies have meant that genome sequence data can be produced faster, easier, and more accurately than ever before. Read lengths that started out at 25 bp on the Solexa/Illumina platform [1] have increased by an order of magnitude in just over half a decade. Such improvements have made possible the creation of ambitious multi-species genome sequencing projects such as Genome 10K (for vertebrates), i5k (for insects), and 959 Nematode Genomes [2-4], among others. A bottleneck for these projects is often the step that needs to convert the raw sequencing data into a high-quality, finished genome sequence. This process of genome assembly is complicated by the different read lengths, read counts, and error profiles that are produced by different NGS technologies. A further challenge is that NGS data for any given genome project sometimes exists as a mixture of reads produced by different technologies.

The need to assemble genomes from NGS data has led to an explosion of novel assembly software. A new generation of assemblers such as EULER [5], ALLPATHS [6], Velvet [7] and ABySS [8] have utilized *de Bruijn* graphs to attack the problem. The *de Bruijn* approach was also used by the SOAPdenovo assembler [9] in generating the first wholly *de novo* assembly of a large eukaryotic genome sequence (the giant panda, *Ailuropoda melanoleuca* [10]). More recent assemblers such as SGA [11] and fermi [12] have capitalized on the increasing length of sequence reads, and utilize string graph approaches, recalling the previous generation of overlap-layout-consensus assemblers. For an overview of these different assembly approaches see [13-16].

Even though *de novo* genome assembly strategies are now capable of tackling the assembly of large vertebrate genomes, the results warrant careful inspection. A comparison of *de novo* assemblies from Han Chinese and Yoruban individuals to the human reference sequence found a range of problems in the *de novo* assemblies [17]. Notably, these assemblies were depleted in segmental duplications and larger repeats leading to assemblies that were shorter than the reference genome. Several recent commentaries that address many of the problems inherent in *de novo* genome assembly [14, 18-22], have also identified a range of solutions to help tackle these issues. These include using complementary sequencing resources to validate assemblies (transcript data, BACs etc), improving the accuracy of insert-size estimation of mate-pair libraries, and trying to combine different assemblies for any genome. There are also a growing number of tools that are designed to help validate existing assemblies, or produce assemblies that try to address specific issues that can arise with *de novo* assemblies. These approaches have included: assemblers that deal with highly repetitive regions [23]; assemblers that use orthologous proteins to improve low quality genome assemblies [24]; and tools that can correct false segmental duplications in existing assemblies [25].

The growing need to objectively benchmark assembly tools has led to several new efforts in this area. Projects such as dnGASP (*de novo* Genome Assembly Project; [26]), GAGE (Genome Assembly Gold-standard Evaluations; [27]), and the Assemblathon [28] have all sought to evaluate the performance of a range of assembly pipelines, using standardized data sets. Both dnGASP and the Assemblathon used simulated genome sequences and simulated Illumina reads, while the GAGE competition used existing Illumina reads from a range of organisms (bacterial, insect, and one human chromosome).

To better reflect the 'real world' usage scenario of genome assemblers, we have organized Assemblathon 2, a genome assembly exercise that uses real sequencing reads from a mixture of NGS technologies. Assemblathon 2 made sequence data available (see Data Description section) for three vertebrate species: a budgerigar (*Melopsittacus undulatus*), a Lake Malawi cichlid (*Maylandia zebra*, also referred to as *Metriaclima zebra*), and a boa constrictor (*Boa constrictor constrictor*). These species were chosen in order to represent a diverse selection of non-mammalian vertebrates, and also because of the availability of suitable sequencing data. For the sake of brevity, these species will henceforth be referred to as simply 'bird', 'fish', and 'snake'. Teams were invited to participate in the contest by submitting assemblies for any or all of these species; in many cases, participating teams were themselves the authors of the assembly tools that they used.

As in the first Assemblathon contest (henceforth, Assemblathon 1) we have attempted to assess the performance of each of each the participating teams by using a variety of metrics. Unlike Assemblathon 1, we do not know what the correct genome sequence should look like for any of the three species. Because of this we make use of various experimental datasets such as Fosmid sequences and optical maps by which to validate the assemblies. A secondary goal of the Assemblathon is to assess the suitability of different metrics by which to assess genome assembly quality, and we employ some novel statistical methods for assessing each assembly

Overall, we find that while many assemblers perform well when looking at a single metric, very few assemblers perform consistently when measured by a set of metrics that assess different aspects of an assembly's quality. Furthermore, we find that assemblers that work well with data from one species may not necessarily work as well with others.

## Data Description

Participating teams (Table 1) had four months in which to assemble genome sequences from a variety of NGS sequence data (Table 2; Additional file 1) that was made available via the Assemblathon website [29]. Each team was allowed to submit one competitive entry for each of the three species (bird, fish, and snake). Additionally, teams were allowed to submit a number of 'evaluation' assemblies for each species. These would be analyzed in the same way as competitive entries, but would not be eligible to be declared as 'winning' entries. Results from the small number of evaluation entries (3, 4 and 0 for bird, fish, and snake respectively) are mostly excluded from the Analyses sections below, but are referenced in the Discussion.

Assemblies were generated using a wide variety of software (Table 1), with greatly varying hardware and time requirements. Details of specific version numbers, software availability, and usage instructions are available for most entries (Tables S2 and S3 in Additional file 2) as are comprehensive assembly instructions (Additional file 3).

Assemblies were excluded from detailed analysis if their total size was less than 25% of the expected genome size for the species in question. Entries from the CoBig$^2$ and PRICE teams did not meet this criterion; their results are included in the Additional file 4 but are not featured in this paper (though see Discussion for information regarding the genic content of the PRICE assembly). Most teams submitted a single file of scaffold sequences, to be split into contigs for contig-based analyses. However, a small number of teams (ABL, CSHL, CTD, and PRICE) submitted one or more entries that consisted only of contig sequences that had not undergone scaffolding.

The submitted assemblies for Assemblathon 2 are available from the Assemblathon website [29] and also from GigaDB [30]. All input reads have been deposited in sequence read archives under the accessions ERP002324 (bird), SRA026860 (fish), and ERP002294 (snake); see Additional file 9 for a detailed list of all associated sequence accessions. Details of the bird sequence data, as well as gene annotations, have also been described separately [31]. The assembled Fosmid sequences for bird and snake that were used to help validate assemblies are also available in GigaDB [32].

Source code for scripts used in the analysis are available from a Github repository [33]. Results for all of the different assembly statistics are available as a spreadsheet (Additional file 4) or as a CSV text file (Additional file 5).

# Additional Files

### Additional file 1

**File format:** Microsoft Word (.docx)
**Title:** Supplementary Data Description
**Description:** Full details of the Illumina, Roche 454, and Pacific Biosciences sequencing data that were made available to participating teams.

### Additional file 2

**File format:** Microsoft Word (.docx)
**Title:** Supplementary Results
**Description:** Additional figures and tables to accompany the main text.

### Additional file 3

**File format:** Microsoft Word (.docx)
**Title:** Assembly Instructions
**Description:** Details provided by participating teams on how to use software to recreate their assemblies. All teams were asked to provide this information.

### Additional file 4

**File format:** Microsoft Excel (.xlsx) spreadsheet
**Title:** Master spreadsheet containing all results
**Description:** Details of 102 different metrics for every assembly. First sheet contains a detailed README explaining all columns. Second sheet contains the data. Third sheet shows z-score values for 10 key metrics for all assemblies. Fourth sheet shows average rankings for all 10 key metrics.

### Additional file 5

**File format:** comma-separated values (.csv)
**Title:** All results
**Description:** This file contains the same information as in sheet 2 of the master spreadsheet (Additional file 4), but in a format more suitable for parsing by computer scripts.

### Additional file 6

**File format:** PDF
**Title:** Bird scaffolds mapped to bird Fosmids
**Description:** Results of using BLAST to align 46 assembled Fosmid sequences to bird scaffold sequences. Each figure represents an assembled Fosmid sequence with tracks showing read coverage, presence of repeats, and alignments to each assembly.

**Additional file 7**

**File format:** PDF
**Title:** Snake scaffolds mapped to snake Fosmids
**Description:** Results of using BLAST to align 24 assembled Fosmid sequences to snake scaffold sequences. Each figure represents an assembled Fosmid sequence with tracks showing read coverage, presence of repeats, and alignments to each assembly.

**Additional file 8**

**File format:** tarred, gzipped archive
**Title:** Bird and snake Validated Fosmid Region (VFR) data
**Description:** The validated regions of the bird and snake Fosmids are available as two FASTA-formatted files. This dataset also includes two FASTA files that represent the 100 nt 'tag' sequences that were extracted from the VFRs.

**Additional file 9**

**File format:** Microsoft Excel (.xlsx) spreadsheet
**Title:** Details of all SRA/ENA/DDBJ accessions for input read data
**Description:** This spreadsheet contains identifiers for all Project, Study, Sample, Experiment, and Run accessions for bird, fish, and snake input read data.

# Analyses

## Statistical description of assemblies

A wide range of basic statistics were calculated for both contigs and scaffold sequences of each assembly (see Additional file 4), including the N50 length. N50 is calculated by summing all sequence lengths, starting with the longest, and observing the length that takes the sum length past 50% of the total assembly length. A related metric, which we adopted for Asssemblathon 1 [28], is the NG50 length. This normalizes for differences in the sizes of the genome assemblies being compared. It is calculated in the same way as N50, except the total assembly size is replaced with the estimated genome size when making the calculation.

The N50 metric is based on using a 50% threshold, but others have sometimes reported this length in combination with other thresholds such as N25 and N75 (e.g.[34]). By extension, if NG values are calculated for all integer thresholds (1–100%), an 'NG graph' can be constructed for all genome assemblies from the same species. The NG graph has several useful properties. First, it allows one to visually compare differences in scaffold lengths for all assemblies. Secondly, the initial data point in any series indicates the size of the longest scaffold for that series. Finally, if a series touches the x-axis (where scaffold NG(X) length = 0), then it indicates that the assembly in question is smaller than the estimated genome size.

Within each species, we observed that assemblies displayed a great deal of variation in their total assembly size, and in their contig and scaffold lengths (Figures 1–3, Figure S1 in Additional file 2, Additional file 4). There is only a modest correlation between scaffold NG50 length and contig NG50 length in bird and snake ($r$ = 0.50 and 0.55 respectively, N.S.), but a stronger correlation in fish ($r$ = 0.78, $P$ < 0.01; Figure S2 in Additional file 2). The snake assemblies from the Phusion and SGA teams have similar scaffold NG50 lengths (3.8 Mbp each) but very different contig NG50 lengths (68 and 25 Kbp respectively). Conversely, the bird assemblies from the MLK and Meraculous teams have similar contig NG50 lengths (36 and 32 Kbp respectively), but extremely different scaffold NG50 lengths (114 and 7,539 Kbp).

When assessing how large each assembly was in relation to the estimated genome size, the MLK bird assembly was observed to be the largest competitive assembly (containing 167% of the 1.2 Gbp estimated amount of sequence). However, a fish evaluation assembly by the IOBUGA team contained almost 2.5 times as much DNA as expected (246% of the estimated 1.0 Gbp). Such large assemblies may represent errors in the assembly process, but they may also represent situations where an assembler has successfully resolved regions of the genome with high heterozygosity into multiple contigs/scaffolds (see Discussion). Among competitive entries, 5 of the 11 bird assemblies were larger than the expected genome size (average ratio = 106.3%; Additional file 4). In contrast fish and snake assemblies tended to be smaller with only 2 out of 10 (fish) and 2 out of 11 (snake) entries larger than the expected genome size (average ratios 92.5% and 96.7% respectively for fish and snake).

Ranking assemblies by their total size or N50/NG50 length can be very misleading if the sequence lengths of the majority of scaffolds are short. In an extreme case, an assembly with the highest N50/NG50 length and largest total size, could comprise of one extremely long scaffold and thousands of very short scaffolds. Following completion of a genome assembly, the primary goal of most genome projects is to find genes, typically using *ab initio* or *de novo* methods of gene prediction [35, 36]. It has been noted that an assembly with a 'gene-sized' scaffold N50 length may be a good target for annotation [37]. More generally, we might consider a 'useful' assembly to be one that has the highest number of scaffolds that are greater than the length of an average gene.

Using 25 Kbp as the approximate length of an average vertebrate gene (see Methods), we calculated what percentage of the estimated genome size in each species consisted of scaffolds that equaled or exceeded this length. This approach suggests that NG50 and N50 can be poor predictors of the suitability of an assembly for gene-finding purposes. For instance, when considering NG50 scaffold length, the Ray bird assembly is the third lowest ranked assembly. However, it comprises 99.2% of the estimated genome size in scaffolds that are at least 25 Kbp (Figure 4). This is not simply because the assembly is larger in size than others (there are four other bird assemblies which are larger in size). Many other assemblies with relatively low NG50 lengths also have high numbers of scaffolds that are over 25 Kbp in length (Figure 4, Figures S3 and S4 in Additional file 2). The snake Curtain assembly has the second lowest NG50 scaffold length (53,529 bp) yet still comprises 80.3% of the estimated genome size in gene-sized length scaffolds. This suggests that someone who is looking to use a genome assembly for gene finding, may not need to be overly concerned by low N50 or NG50 values.

### Presence of core genes

The completeness and correctness of genic sequences in assemblies is of paramount importance for diverse applications. For many assembled genomes, transcriptome data has been acquired in parallel, and such data could be mapped back to the assemblies to directly assess the presence of genes. However for the three species in this study, very little full-length cDNA or RefSeq data were available (Table S3 in Additional file 2).

Therefore we restricted our attention to measuring the presence of highly conserved genes that should be present in nearly all eukaryotic genomes and which should be highly similar to orthologs in 'known' genomes. For this purpose we used a set of 458 'core eukaryotic genes' (CEGs) [38], and assessed their presence by testing for 70% or greater presence of each gene within a single scaffold, as compared to an HMM (hidden Markov model) for the gene. This analysis was carried out using CEGMA ([38], Methods). The analysis could thus assess presence, but not accuracy of the given genes within the assemblies. However, CEGMA outputs a predicted protein sequence for each gene, and we note that for a given species, for a given gene, the protein sequences derived from different assemblies capturing the gene were largely identical, suggesting that most genes were 100% present and accurate (Figure S5 in Additional file 2). Differences between captured genes could be attributable to polymorphism, assembly defects, or limitations of CEGMA in distinguishing between paralogous genes.

Nearly all of the 458 CEGs were found in at least one assembly (442, 455, 454 for bird, fish, and snake; CEGMA gene predictions have been submitted to GigaDB [39]). We evaluated the assemblies by computing the fraction of these totals that were present in a given assembly, finding in nearly all cases that these fractions varied from 85–95%, a significant variation in utility (Figure 5, Tables S4-S6 in Additional file 2). Differences in performance could be attributable to several reasons, including fracturing of a given genic region across multiple scaffolds within an assembly, or exons lying in gaps within a single scaffold. It is also possible that for some, highly paralogous, genes CEGMA is detecting a paralog and not the true ortholog.

To address these issues we inspected the secondary output of CEGMA which reports statistics for a published subset of core genes[40]. The 248 CEGs in this subset correspond to the most highly conserved, least paralogous of the original set of 458 CEGs. For this subset, CEGMA also reports on how many partial matches it found (see Methods). The results from using either set of CEGs are highly correlated (Figure S6 in Additional file 2) but reveal that many assemblies contain additional core genes that are too fragmented to be detected by the original analysis (Figure S7 in Additional file 2).

### Analysis of Fosmid sequences in bird and snake

Fosmid sequence data were made available for bird and snake (see Methods) and these sequences were used to help assess the accuracy of the respective genome assemblies. The assembled Fosmid sequences (46 for bird and 24 for snake) were first assessed for their read coverage and repeat content (see Methods), and were then aligned to scaffolds from each assembly. This analysis revealed a great deal of variety in the repeat content and read coverage of different Fosmids, and also in the number of different assemblies that would align to an individual Fosmid sequence. Most Fosmids were well represented by many assemblies, with minor gaps in the alignments to scaffolds corresponding to breaks in Fosmid read coverage (Figure 6A). Other Fosmids with higher repeat content were not so well represented in the assemblies; as might be expected, repeats from the Fosmids were often present in multiple scaffolds in some assemblies (Figure 6B). Details of alignments, coverage, and repeats for all 90 Fosmids are available in Additional files 6 and 7.

It is possible that some of the assembled Fosmid scaffold sequences are themselves the result of incorrect assemblies of the raw Fosmid read data. If the Fosmids are to be used to help validate assemblies, only those regions that we believe to be accurately assembled should be utilized. To ensure this, we extracted just the Fosmid regions that are supported by a correct alignment (of 1 Kbp minimum length) to one or more scaffold sequences from any of the assemblies. This produced a number of validated Fosmid regions (VFRs) from the Fosmids (86 for bird and 56 for snake; Additional file 8). These VFRs covered the majority of the total Fosmid length in both species (bird: ~99% coverage, 1,035 out of 1,042 Kbp; snake: ~89% coverage, 378 out of 422 Kbp; see Methods). In most cases, the regions of Fosmids that were not validated coincided with low read coverage and/or the presence of repeats (see Supplementary

Material). VFRs were then used as trusted reference sequences for various analyses by which to assess the accuracy of the assemblies.

**COMPASS analysis of VFRs**

When a reference sequence is available, genome assemblies can be assessed by how much of the reference is covered by alignments of scaffolds to the reference. A higher fractional coverage of the total reference sequence length is generally preferred to lower coverage. However, a reference that has high coverage doesn't reveal how much sequence was needed to achieve that level of coverage, or how much duplication there was among different scaffolds that aligned to the reference. To address these limitations of using coverage alone, we propose three new quantities that we define as validity, multiplicity, and parsimony. All of these metrics can be calculated from the alignment of an assembled sequence to a (possibly partial) trusted reference sequence. A tool, COMPASS [33, 41] was used to calculate these metrics.

To explain these metrics, let us define four length sets (Figure 7). The first is the set of assembled scaffold lengths, ($S_i$); next, a set of reference sequence lengths ($R_i$); then the lengths of the alignments of scaffolds to the reference ($A_i$); and finally the lengths of the 'coverage islands' ($C_i$), which consist of ranges of continuous coverage (on a reference sequence) by one or more alignment(s) from the assembly. Fractional *coverage* of the reference is then found by $\Sigma C_i/\Sigma R_i$. We define *validity* to be $\Sigma A_i/\Sigma S_i$, which reflects the alignable, or validatable fraction of assembled sequence. We define *multiplicity* as $\Sigma A_i/\Sigma C_i$, which reflects the ratio of the length of alignable assembled sequence to covered sequence on the reference; higher multiplicity implies expansion of repeats in the assembly, lower multiplicity implies repeat collapse. Finally, we define *parsimony* as multiplicity divided by validity, or $\Sigma S_i/\Sigma C_i$. This final metric can be thought of as the "cost" of the assembly: how many bases of assembled sequence need to be inspected, in order to find one base of real, validatable sequence. The alignment procedure used to determine these four metrics should be tuned according to the nature of the reference sequence. I.e. ungapped alignments can be considered if the reference was generated from the same haploid sample as the assembly, and less stringent alignments can be considered if greater divergence is expected between the assembly and reference. A comparison of two or more assemblies may reveal similar levels of coverage, but differing levels of validity, multiplicity, and parsimony. This is especially the case if one assembly is smaller than another (leading to higher validity of the smaller assembly), or if one assembly contains more duplications (leading to higher multiplicity).

COMPASS statistics were calculated for assemblies from bird (Figure 8) and snake (Figure 9). Fosmid coverage was seen to vary between assemblies, particularly in snake, but was not correlated with genome assembly size (Figure S8 in Additional file 2), reinforcing the notion that overall assembly size is not necessarily a good predictor of assembly quality.

The results suggest that the bird assembly by the Newbler-454 team performed very well, with the highest levels of coverage and validity, and lowest values for multiplicity and parsimony among all competitive bird assemblies. The Allpaths assembly was the only other competitive entry to rank in the top five assemblies for all COMPASS metrics. For snake, assemblies by the Ray, BCM-HGSC, CRACS, and SGA teams all scored well with high values of coverage and validity. The Ray assembly was ranked 1st overall, and also ranked 1st for all individual measures except multiplicity (where it still had a better than average performance).

In most cases, coverage and validity are highly correlated in both species ($r$ = 0.70 and $r$ = 0.94 for bird and snake respectively). One notable exception to this was the MLK bird assembly which ranked 3rd for coverage but 10th for validity. This makes sense in light of the fact that this was a very large assembly (equivalent to 167% of the expected genome size; see Additional file 4) which also had the highest multiplicity of any assembly from either species. Inclusion of extra, non-alignable sequence in an assembly and/or expansion of repetitive sequence can both contribute to high parsimony values (the former decreases validity, the latter increases multiplicity). However, the true copy number of any repeat sequence can be difficult to ascertain, even in very high quality assemblies; this can make comparisons of multiplicity difficult to evaluate.

The COMPASS program also produces cumulative length plots (CLPs) that display the full distribution of a set of sequence lengths. These were calculated for all competitive assemblies using the set of scaffold lengths (Figure 10) and the set of alignment lengths of the scaffolds to the VFRs (Figure 11). Unlike the single-value metrics of coverage, validity, multiplicity, and parsimony, CLPs allow for comparisons across the full spectrum of scaffold or alignment lengths, and can reveal contrasting patterns. For instance, while the bird scaffold length plots are widely separated, the alignment length plots cluster more tightly, suggesting that scaffolding performance could vary more than the performance of producing contigs. Among the snake assemblies, there is an intriguing contrast in the performance of the Ray assembly; it is comparatively poor in terms of its scaffold CLPs, but it outperforms all other assemblies based on its alignment CLPs.

**Assessing short-range accuracy in validated Fosmid regions**

VFR data was also used to assess the short-range accuracy of contig and scaffold sequences from the bird and snake assemblies. Each VFR sequence was first divided into non-overlapping 1,000 nt fragments (988 fragments for bird and 350 for snake; Additional file 8). Pairs of short (100 nt) 'tag' sequences from the ends of each fragment were then extracted in order to assess:

a) How many tags mapped anywhere in the assembly
b) How many tags matched uniquely to one contig/scaffold
c) How many pairs of tags matched the same contig/scaffold (allowing matches to other sequences)
d) How many pairs of tags matched uniquely to the same contig/scaffold
e) How many pairs of tags matched uniquely to the same contig/scaffold at the expected distance apart (900 ± 2 nt, to allow for short indels in the assembly and/or Fosmids)

Failure to map uniquely to a single contig or scaffold sequence might be expected when VFR tag sequences are derived from repetitive regions of the Fosmids, or if a Fosmid was incompletely assembled. To address this, a final summary statistic by which to evaluate the assemblies was calculated. This summary score is the product of the number of tag pairs that matched the same contig/scaffold (uniquely or otherwise) and the percentage of the uniquely matched tag pairs that map at the expected distance (i.e. $c * (e / d)$ ). This measure rewards assemblies that have produced sequences that contain both tags from a pair (at any distance) and which have a high proportion of uniquely mapped tag pairs that are mapped the correct distance apart.

Overall, assemblies were broadly comparable when being assessed by this metric and produced similar summary scores for contigs (Additional file 4) and scaffolds (Figure 12, Tables S5 and S6 in Additional file 2). In bird, the CBCB assembly had the 8th lowest accuracy (91.8%) for placing uniquely mapped tag pairs at the correct distance apart in a scaffold, but had the highest number of tag pairs that mapped correctly to the same scaffold (910 out of 988). This helped contribute to the highest overall VFR tag summary score. In snake, the ABYSS assembly produced the highest summary score, but this was only slightly higher than the 2nd-placed MERAC assembly (305.8 vs 305.0). The vast majority of tags that were mapped at incorrect distances, were usually mapped within 10 nt of the correct distance. Such mismappings might reflect instances of small-indel heterozygosity in the underlying genomes. However, a small number of mismappings were over much great distances (Tables S5 and S6 in Additional file 2).

## Optical map analysis

The Optical Mapping System [42, 43] creates restriction maps from individual genomic DNA molecules that are assembled, *de novo*, into physical maps spanning entire genomes. Such maps have been successfully applied to many large-scale sequencing projects[44-48] and to the discernment of human structural variation [49]. Recent work has centered on approaches that integrate map and sequence data at an early stage of the assembly process [50].

Optical maps were constructed for all three species and were used to validate the long- and short-range accuracy of the scaffold sequences (see Methods). Because the mapping process requires scaffolds to be at least 300 Kbp in length, with nine restriction sites present, a number of assemblies had no sequence that could be used. In contrast, some assemblies could use nearly all of their input sequence (e.g. 95.8% of the CSHL fish assembly was used, see Additional file 4).

The optical map results describe two categories of global alignments, either 'restrictive' (level 1) or 'permissive' (level 2), and one category of local alignment (level 3). High coverage at level 1 suggests that the optical map and the scaffold sequences are concordant. Level 2 coverage reveals cases where there are minor problems in the scaffolds, and coverage at level 3 represents regions of the scaffolds that may reflect bad joins or chimeric sequences.

The results for bird (Figure 13) show many assemblies with high amounts of level 1 coverage, with relatively small differences in the total amount of alignable sequence. The bird SGA assembly was notable for having the second highest amount of level 1 coverage, but ranked 8th overall due to fewer alignments at levels 2 and 3. Among the assemblies that could be subjected to optical map analysis, the MLK bird assembly ranked last in terms of the total length of usable scaffold sequence. However, it ranks 2nd based on the percentage of input sequence that can be aligned to the optical map (see Additional file 4).

For the fish assemblies, the optical mapping results show very low proportions of alignable sequence (Figure 14), and alignable sequence is predominantly in the level 3 category, reflecting local alignments. As with bird, ranking assemblies by their total length of alignable sequence may not paint the most accurate picture of how the assemblies perform. The SGA assembly, which ranks 7th overall by this measure, had the most level 1 coverage of any assembly, and the Ray assembly, which ranked 8th by this measure, had the highest proportion of input sequence that was alignable.

The results for the snake assemblies (Figure 15) were somewhat intermediate to those for bird and fish. Many snake assemblies showed patterns of alignment to the optical map that were predominantly in the level 2 coverage category. The SOAPdenovo assembly had the most input sequence (1.5 Gbp) of any assembly (across all three species) that was suitable for alignment to the optical map. However, this assembly ranked third last with only 13.6% of the sequence being aligned at any coverage level (see Discussion and Additional file 3 for reasons which might have caused this result).

### REAPR analysis

REAPR [51] is a software tool that analyses the quality of an assembly, using the information in remapped paired-end reads to produce a range of metrics at each base of the assembly. It uses Illumina reads from a short fragment library to measure local errors such as SNPs or small insertions or deletions, by considering perfect and uniquely mapped read pairs. Reads from a large fragment library are used to locate structural errors such as translocations. Each base of an assembly is analysed and designated as "error-free" if no errors are detected by both the short and long fragment library reads. The large fragment size reads are also used to detect incorrect scaffolds, thereby generating a new assembly by breaking at incorrect gaps in the original assembly.

An overall summary score was generated for each assembly by combining the number of error-free bases with the scaffold N50 length, calculated before and after breaking the assembly at errors, as follows:

$$\text{Number of error free bases} * (\text{broken N50})^2 / (\text{original N50})$$

Normalization is also applied within each species (see Methods). The summary score rewards local accuracy, overall contiguity and correct scaffolding of an assembly.

The REAPR summary scores reveal large differences between the quality of different assemblies, and show that scores are higher in snake than in bird or fish (Figure 16; Additional file 4). A detailed inspection of the REAPR results suggests that the fish genome is highly repetitive, with many collapsed repeats called, compared with the assemblies of snake and bird. There is an overall trend showing the expected trade-off between accuracy and contiguity. For example, the Ray snake assembly is very conservative with modest N50 scaffold lengths of 132 Kbp and 123 Kbp (before and after breaking by REAPR). This is in contrast to the snake Curtain assembly that has an N50 length of 1,149 Kbp which is reduced to 556 Kbp after breaking. Since the REAPR score is rewarding correct scaffolding, it is not simply correlated with scaffold N50 length. For example, the snake Meraculous and SGA assemblies have comparable numbers of error-free bases called, but the N50 before and after breaking makes the difference between their summary scores. Although the Meraculous assembly had lower N50 values than those of the SGA assembly, it made proportionally fewer errors, so that it was ranked above SGA.

REAPR did not utilize all available sequence data when evaluating assemblies in each species. This was particularly true for the bird data, where a large number of libraries were available to the participating teams (see Methods). Assemblies that were optimized to work with sequences from a particular library may have been penalized if REAPR used sequences from a different library to evaluate the assembly quality.

**Ranking assemblies**

This paper describes many different metrics by which to grade a genome assembly (see Additional file 4 for a complete list of all results). Clearly many of these are interdependent (e.g. N50 and NG50) and to a degree, every metric is a compromise between resolution and comprehensibility. However, many of the metrics used in this study represent approaches that are largely orthogonal to each other. In order to produce an overall assessment of assembly quality, we chose ten 'key' metrics from which to calculate a final ranking. These are as follows:

1. NG50 scaffold length: a measure of average scaffold length that is comparable between different assemblies (higher = better).
2. NG50 contig length: a measure of average contig length (higher = better)
3. Amount of scaffold sequence gene-sized scaffolds (>= 25 Kbp): measured as the absolute difference from expected genome size, this helps describe the suitability of an assembly for gene finding purposes (lower = better).
4. CEGMA, number of 458 core genes mapped: indicative of how many genes might be present in assembly (higher = better).
5. Fosmid coverage: calculated using the COMPASS tool, reflects how much of the VFRs were captured by the assemblies (higher = better).
6. Fosmid validity: calculated using the COMPASS tool, measures the amount of the assembly that could be validated by the VFRs.

7. VFR tag scaffold summary score: number of VFR tag pairs that both match the same scaffold multiplied by the percentage of uniquely mapping tag pairs that map at the correct distance apart. Rewards short-range accuracy (higher = better).
8. Optical map data, Level 1 coverage: a long-range indicator of global assembly accuracy (higher = better).
9. Optical map data, Levels 1+2+3 coverage: indicates how much of an assembly correctly aligned to an optical map, even if due to chimeric scaffolds (higher = better).
10. REAPR summary score: rewards short- and long-range accuracy, as well as low error rates (higher = better).

For the fish assemblies, the lack of Fosmid sequence data meant that we could only use seven of these key metrics. In addition to ranking assemblies by each of these metrics and then calculating an average rank (Figures S9–S11 in Additional file 2), we also calculated z-scores for each key metric and summed these. This has the benefit of rewarding/penalizing those assemblies with exceptionally high/low scores in any one metric. One way of addressing the reliability and contribution of each of these key metrics is to remove each metric in turn and recalculate the z-score. This can be used to produce error bars for the final z-score, showing the minimum and maximum z-score that might have occurred if we had used any combination of nine (bird and snake) or six (fish) metrics.

The results for the overall rankings of the bird, fish, and snake assemblies reveal a wide range of performance within, and between, species (Figures 17–19). See the Discussion for a species-by-species analysis of the significance of these z-score based rankings.

**Analysis of key metrics**

After using the key metrics to rank all of the genome assemblies, we can ask how different is this ordering compared to if we had only ranked the assemblies by the widely-used measure of N50 scaffold length. All three species showed strong correlations between N50 and the final z-score for each assembly (Figure 20). For fish and snake, which showed significant correlations ($P < 0.01$), the highest N50 length also occurred in the highest ranked assembly.

More generally, we can look to see how correlated the 10 key metrics are across all of the assemblies (including evaluation entries). We illustrate this by means of species-specific correlation plots (Figures S12–S14 in Additional file 2). Although we observe strong, assembly-specific correlations between various metrics, many of these are not shared between different assemblies. This suggests that it is difficult to generalize from one assembly problem to another. However, when pooling data from bird and snake — the two species for which all 10 key metrics were available — we find a small number of significant correlations between certain key metrics (Figure S15 in Additional file 2). Across these two species, we observe significant correlations between the two COMPASS metrics of coverage and validity ($r = 0.84$, $P < 0.0001$), between the two optical map coverage metrics ($r = 0.85$, $P < 0.0001$) and between the VFR scaffold summary score and the number of scaffolds that are >= 25 Kbp ($r = 0.88$, $P < 0.0001$). In fish, where optical map data were available, we did not find a significant correlation between the two optical map coverage metrics ($r$ = 0.21; Figure S13 in Additional file 2). Furthermore, the bird assembly with the highest level 1–3 coverage of the optical maps (SOAP**), ranked only 9th in its level 1 coverage.

We visualized the performance of all assemblies across all key metrics by means of a heat-map (Figure S16 in Additional file 2) and a parallel coordinate mosaic plot (Figure 21). Comparing the assemblies in this way reveals that there are clear weaker outliers for each species, and that there are few assemblies which are particularly strong across all key metrics.

# Discussion

## Overview

Using a mixture of experimental data and statistical approaches, we have evaluated a large number of *de novo* genome assemblies for three vertebrate species. It is worth highlighting the fact that the true genome sequences for these species remain unknown and that this study is focused on a *comparative* analysis of the various assemblies.

## Interspecific vs intraspecific variation in assembly quality

Overall, we observed that bird assemblies tended to have much longer contigs (Figure S1 in Additional file 2), longer scaffolds (Figures 1–3), and had more assemblies that comprised 100% (or more) of the estimated genome size (1.2 Gbp for bird) than the other two species. This is potentially a reflection of the much higher coverage of the bird genome than the other two species (Table 2). Bird assemblies also performed better than fish and snake when assessed by the optical map data (Figures 13–15). The optical map data also suggested that fish assemblies were notably poorer than the other two species. These widely varying results suggest that differing properties of the three genomes pose different challenges for assemblers, but it may also reflect differences in the qualities of the three optical maps.

Several other metrics suggest that it is the snake genome that, on average, had the highest scoring assemblies of any of the species. For example, the average number of core eukaryotic genes (CEGs) detected in the competitive assemblies of bird, fish, and snake was 383, 418 and 424 respectively (Additional file 4). The fish assemblies tended to be the lowest quality of the three species with the REAPR analysis suggesting that only 68% of the bases across all competitive fish assemblies are error free compared to 73% and 75% of the bird and snake assemblies (Additional file 4).

Genome size alone does not seem to be a factor in the relative increases in quality of the snake — and to a lesser extent, bird — assemblies relative to that of fish; the snake (boa constrictor) is estimated to have the largest genome of the three species (1.6 Gbp vs 1.2 Gbp for bird and 1.0 Gbp for fish). The most likely factors that account for these interspecific differences would be the differing levels of heterozygosity and/or repeat content in the three genomes. In agreement with this hypothesis, Lake Malawi cichlids are known to be highly genetically diverse, resulting from extensive hybridization that leads to high levels of heterozygosity [52, 53]. The REAPR analysis also suggests that repeats in the fish genome might pose more of an issue for assemblers than the bird and snake genomes. The extent to which repeat content and heterozygosity made it harder to assemble the bird and snake genomes is less clear. The RepeatMasker analysis of the Fosmid sequences revealed there to be more repeats in snake than bird (13.2% of total Fosmid sequence length vs 8.6%). Observed heterozygosity for boa constrictors has been shown to be in the range of 0.36–0.42 [54] whereas a much wider range (0.1–0.8) has been described for budgerigars (Zhang and Jarvis, personal communication).

Aside from possible differences in the genomes of the three species, it should also be noted that there are interspecific differences with regards to the sequencing data that was available for each species. E.g. all of the short-insert Illumina libraries in fish had overlapping paired reads, whereas in snake they were all non-overlapping (see Supplementary Methods). These differences mean that assemblers that were well-suited for working with the fish data may not be as suited for working with the snake data, and vice versa. The bird genome was different to the other species in having many more libraries available with many different insert sizes (see Additional file 1) and also in having sequencing data available from three different platforms (discussed below, see 'The effects of combining different data types').

### Bird assembly overview

For the budgerigar genome, we found the BCM-HGSC assembly to be the highest ranked competitive assembly when using the sum z-score approach (Figure 17). However, if evaluation assemblies are included then the BCM-HGSC evaluation entry (BCM*) produces a notably higher sum z-score; the reasons for the differences between these two BCM-HGSC assemblies are discussed below (see 'The effects of combining different data types'). The two evaluation assemblies from the SOAPdenovo team (SOAP* and SOAP**) also rank higher than the competitive SOAP entry.

Among the competitive assemblies, the Allpaths and Newbler entries rank closely behind the assembly from the BCM-HGSC team in terms of overall z-score. While the competitive BCM-HGSC entry performs well across many of the key metrics, it would not be ranked 1st overall if any one of three different key metrics (CEGMA, scaffold NG50, and contig NG50) had been excluded from the calculation of the z-score. The overall heterogeneity of the bird assemblies is further underlined by the fact that 6 of the 14 entries (including evaluation assemblies) would rank 1st (or joint 1st) when ranked separately by each of the ten key metrics.

Ordering the assemblies by their average rank rather than by z-score produces a slightly different result (Figure S9 in Additional file 2). Assemblies from BCM-HGSC and Allpaths switch 1st and 2nd places, but both are still placed behind the BCM* assembly. The CBCB entry ranks higher using this method, moving to 3rd place among competitive entries.

It should be noted that the top three-ranked competitive bird assemblies each used a very different combination of sequencing data: BCM-HGSC used Illumina + Roche 454 + PacBio, Allpaths only used Illumina, and Newbler only used Roche 454. Therefore, the similarity in overall assembly rankings should be weighed against the different costs that each strategy would require.

### Fish assembly overview

The lack of Fosmid sequence data for the Lake Malawi cichlid removed three of the key metrics that were used for the other two species. Overall, the fish assemblies could broadly be divided into three groups with the first group consisting of the most highly ranked assemblies generated by the teams BCM-HGSC, CSHL, Symbiose and Allpaths (Figure 18). The BCM-HGSC assembly scored highly in most key metrics, and excelled in the measure of scaffold N50 length. This is the only key metric which, if excluded, would remove the BCM-HGSC assembly from 1st place when using a z-score ranking system. An ordering of assemblies based on their average rank provides only minor differences to that of the z-score ranking (Figure S10 in Additional file 2).

The CSHL team submitted two additional evaluation assemblies for fish. Their competitive assembly ranked 2nd overall, and was produced by the metassembler tool [55] which combined the results of two separate assemblies (CSHL* and CSHL**, that were made using the Allpaths and SOAPdenovo assemblers respectively). Both of these CSHL evaluation assemblies were produced using the default parameters for the assembly software in question (though for the SOAPdenovo assembly, CSHL team also used Quake [56] to error correct the reads before assembly). The CSHL Allpaths assembly ranked slightly higher than the competitive entry from the Allpaths team, though this is only apparent in the z-score rankings (they produce the same average rank, Figure S10 in Additional file 2). In contrast, the CSHL SOAPdenovo entry ranked much lower than the evaluation assembly from the SOAPdenovo team entry (SOAP*).

Snake assembly overview

The snake assemblies provided the clearest situation where one competitive assembly outperformed all of the others. The SGA assembly scored highly in eight of the ten key metrics, producing a final z-score that was notably higher than that of the Phusion assembly that ranked 2nd (Figure 19). If any one of the ten key metrics were removed from the analysis, the SGA assembly would still rank 1st by either ranking method. Ordering the assemblies by their average rank produced a near identical ranking when compared to using z-scores (Figure S11 in Additional file 2).

It should be noted that the SOAPdenovo entry was generated at a time when some of the Illumina mate-pair libraries were temporarily mislabelled in the data instruction file (details of 4 Kbp and 10 Kbp libraries were mistakenly switched). The fact that their assembly was produced with incorrectly labelled data was not noticed until all assemblies had been evaluated, and this may therefore have unfairly penalized their entry. A corrected assembly has subsequently been made available [57] which provides a ~6-fold increase in the scaffold NG50 length compared to the original entry.

Despite the apparent pre-eminence of the SGA assembler it should still be noted that this assembly only ranked 1st in one of the ten key metrics that was used, and ranked 7th in another (the amount of gene-sized scaffolds). Furthermore, seven different assemblies would rank 1st if assessed by individual metrics from the set of ten. This reinforces the challenges of assessing the overall quality of a genome assembly when using multiple metrics.

## Assembler performance across all three species

SGA, BCM-HGSC, Meraculous, and Ray were the only teams to provide competitive assemblies for all three species (SOAPdenovo provided entries for all species, but only included an evaluation assembly for fish). However, other teams included assemblies for at least two of the species so it is possible to ask how many times did an assembler rank 1st for any of the key metrics that were evaluated. Theoretically, an assembler could be ranked 1st in 27 different key metrics (ten each for bird and snake, and seven for fish).

Excluding the evaluation entries, we observed that assemblies produced by the BCM-HGSC team achieved more 1st place rankings (five) than any other team. Behind the BCM-HGSC team were Meraculous and Symbiose (four 1st place rankings each), and the Ray team (three 1st place rankings). The meraculous assembler was notably consistent in its performance in the same metric across different species, ranking 1st, 2nd, and 1st in the level 1 coverage of the optical maps (for bird, fish, and snake respectively). The result for Ray is somewhat surprising as the three Ray assemblies only ranked 7th, 7th and 9th overall among competitive entries (for bird, fish, and snake respectively).

These analyses reveal that — at least in this competition — it is very hard to make an assembly that performs consistently when assessed by different metrics within a species, or when assessed by the same metrics in different species.

## The effects of combining different data types in bird

For the bird genome, we provided three different types of sequencing data: Illumina, Roche 454, and Pacific Biosciences (PacBio). However, only four teams attempted to combine sequence data from these different platforms in their final assemblies.

The BCM-HGSC team used all three types of sequence data in their competitive entry (BCM), but did not use the PacBio data in their evaluation entry (BCM*). For their competitive assembly, PacBio data were used to fill scaffolding gaps (runs of Ns) but otherwise this assembly was generated in the same way as the evaluation entry. Although gap-filling in this manner led to longer contigs (Figure S1 in Additional file 2), the overall effect was to produce a lower-ranked assembly (Figure 17). This is because inclusion of the higher error-rate PacBio data led to a marked decrease in the coverage and validity measures produced by COMPASS. This, in turn, was because the Lastz tool [58] that was used for alignment was run with a zero penalty for ambiguous characters (Ns), rather than the default penalty score. Consequently, errors in PacBio sequence used in the scaffolding gaps caused breaks in alignments and exclusion of shorter alignments between gaps. If this setting were changed to penalize matches to ambiguous bases in the same way as mismatched unambiguous bases, then it would likely reverse the rankings of these two assemblies.

In addition to a competitive entry, the SOAPdenovo team included two evaluation assemblies for bird which both ranked higher than their competitive entry. The evaluation entries differed in using only Illumina (SOAP*) or Illumina plus Roche 454 (SOAP**) data. Inclusion of the Roche 454 data contributed to a markedly better assembly (Figure 18), but again this was mostly achieved through increased coverage and validity when compared to the SOAP* assembly.

The other teams that combined sequencing data were the CBCB team that used all three types of data and the ABL team which used Illumina plus Roche 454 data. Both of these teams only submitted one assembly so it is not possible to accurately evaluate the effect of combining data sets compared to an assembly which only used one set of sequence data. The CBCB team have separately reported on the generation of their entry, as well as additional budgerigar assemblies, and have described the effects of correcting PacBio reads using shorter Illumina and 454 reads[59]. Their assembly performed competently when assessed by most metrics but was penalized by much lower NG50 scaffold lengths compared to other bird assemblies. It should also be noted that the Ray assembler [60] that was used for the Ray fish assembly, was designed to work with Illumina and Roche 454 reads, but this team chose to only use the Illumina data in their assembly.

Overall, the bird assemblies that attempted to combine multiple types of sequencing data ranked 1st, 2nd, 5th, 7th, and 14th when assessed by all key metrics. The two assemblies that included PacBio data (BCM and CBCB) had the highest and second-highest contig NG50 lengths among all competitive bird assemblies (Figure S1 in Additional file 2), suggesting that inclusion of PacBio data may be particularly useful in this regard. However, it may be desirable to correct the PacBio reads using other sequencing data as was done by the CBCB team, a process that may have been responsible for the higher values of coverage and validity in this assembly compared to the BCM-HGSC entry.

Aside from differences in assembly quality, it should also be noted that the generation of raw sequence data from multiple platforms will typically lead to an increase in sequencing costs. This was not an aspect factored into this evaluation, but should be an important consideration for those considering mixing different data types. It should also be pointed out that not all assemblers are designed to work with data from multiple sequencing platforms.

### Size isn't everything

Assemblies varied considerably in size, with some being much bigger or smaller than the estimated genome size for the species in question. However, very large or small assemblies may still rank highly across many key metrics. E.g. among competitive entries, the Ray team generated the smallest fish assembly (~80% of estimated genome size), but this ranked 3rd overall (Additional file 4). The PRICE snake assembly was excluded from detailed analysis because it accounted for less than 25% of the estimated snake genome size. This team used their own assembler[61] and implemented a different strategy to that used by other teams, focusing only on assembling the likely genic regions of the snake genome. They did this by looking for matches from the input read data to the gene annotations from the green lizard

(*Anolis carolinensis*); this being the closest species to snake that has a full set of genome annotations. While their assembly only comprises ~10% of the estimated genome size for the snake, it contains almost three-quarters (332 out of 438) of the core eukaryotic genes that are present across all snake assemblies (see Additional file 4). While this is still fewer than any other snake assembly, it would be ranked highest if evaluating assemblies in terms of 'number of core genes per Mbp of assembly (Figure S17).

**Lessons learned from Assemblathon 2**

The clear take-home message from this exercise is the lack of consistency between assemblies in terms of interspecific as well as intraspecific comparisons. An assembler may produce an excellent assembly when judged by one approach, but a much poorer assembly when judged by another. The SGA snake assembly ranked 1st overall, but only ranked 1st in one individual key metric, and ranked 5th and 7th in others. Even when an assembler performs well across a range of metrics in one species, it is no guarantee that this assembler will work as well with a different genome. The BCM-HGSC team produced the top ranking assembly for bird and fish but a much lower-ranked assembly for snake. Comparisons between the performance of the same assembler in different species are confounded by the different nature of the input sequence data that was provided for each species.

By many metrics, the best assemblies that were produced were for the snake, a species that had a larger genome than the other two species, but which had fewer repeats than the bird genome (as assessed by RepeatMasker analysis). The snake dataset also had the lowest read coverage of all three species, with less than half the coverage of the bird (Table 2). Higher levels of heterozygosity in the two other genomes are likely to be responsible for these differences.

We used ten 'key metrics' which each capture a slightly different facet of assembly quality. It is apparent that using a slightly different set of metrics could have produced a very different ranking for many of the assemblies (Figures 17–19, Figures S9–S11 in Additional file 2). Two of these key metrics are based on alignments of scaffolds to optical maps and these metrics sometimes revealed very different pictures of assembly quality. E.g. the SGA fish assembly had very high level 1 coverage of the optical map, reflecting global alignments that indicate scaffolds lacking assembly problems. In contrast, this assembly ranked below average for the total coverage (levels 1–3) of the optical maps. This suggests that many other assemblies were better at producing shorter regions of scaffolds that were accurate, even if those scaffolds were chimeric.

N50 scaffold length — a measure that came to prominence in the analysis of the draft human genome sequence [62] — remains a popular metric. Although it was designed to measure the contiguity of an assembly, it is frequently used as a proxy by which to gauge the quality of a genome assembly. Continued reliance on this measure has attracted criticism (e.g. [15]) and others have proposed alternative metrics such as 'normalized N50' [63] to address some of the criticisms. As in Assemblathon 1 [28], we find that N50 remains highly correlated with our overall rankings (Figure 20). However, it may be misleading to rely solely on this metric when assessing an assembly's quality. E.g. the SOAP bird assembly has the 2nd highest N50 length but ranked 6th among competitive assemblies based on the overall z-score. Conversely, assemblies with low scaffold N50 lengths may excel in one or more specific measures of assembly quality. E.g. the Ray snake assembly ranked 9th for N50 scaffold length but ranked 1st in the two COMPASS metrics of coverage and validity.

Recently, another assembly quality metric has been proposed that uses alignments of paired-end and mate-pair reads to an assembly to generate Feature-Response Curves (FRC)[15, 64]. This approach attempts to capture a trade-off between accuracy and continuity, and has recently been used to assess a number of publicly available genome assembly datasets including the snake assemblies that were submitted for Assemblathon 2 [65]. The authors used the read alignments to generate a number of features which can be evaluated separately or combined for an overall view of assembly accuracy. They identified SGA and Meraculous as producing the highest ranking assemblies, results which agree with our findings (SGA and Meraculous ranked 1st and 3rd). They also echoed our conclusions that focusing on individual metrics can often produce different rankings for assemblers.

Combining multiple assemblies from different assembly pipelines in order to produce an improved assembly was an approach used in the assembly of the rhesus macaque genome[66]. It might therefore be expected that an improved assembly could be made for each of the three species in this study. The results from the CEGMA analysis (Figure 5) indicate that this may be possible, at least in terms of the genic content of an assembly. Three fish assemblies (CSHL, CSHL*, and SOAP*) were all found to contain the most core genes (436 out of 458 CEGs), but 455 CEGs were present across all assemblies. Combining assemblies is the approach that the GAM team used for their snake assembly. The GAM program (Genomic Assemblies Merger)[67] combined separate assemblies produced by the CLC and AbySS assemblers [8, 68]. However, the resulting assembly scored poorly in most metrics. In contrast, the metassembler entry from the CSHL team produced a high-ranking assembly, but one that was only marginally better than the two source assemblies that it was based on.

One important limitation of this study is that we did not assess the degree to which different assemblers resolved heterozygous regions of the genome into separate haplotypes. Therefore we do not know whether the larger-than-expected assemblies may simply reflect situations where an assembler successfully resolved a highly heterozygous region into two separate contigs. Some assemblers are known to combine such contigs into one scaffold where the heterozygous region appears as a spurious segmental duplication [25]. Many assemblers only produce only a haploid consensus version of a target diploid genome. This is partly a limitation of the FASTA file format and a current effort to propose a new assembly file format is ongoing. This FASTG format [69] is intended to allow representation of heterozygous regions (and other uncertainties) and could lead to more accurate assessments of genome assembly quality in future.

A final, but important, point to note is that many of the assemblies entered into this competition were submitted by the authors of the software that was used to create the assembly. These entries might therefore be considered to represent the best possible assemblies that could be created with these tools; third-party users may not be able to produce as good results without first gaining considerable familiarity with the software. Related to this point are the issues of: 'ease of installation', 'quality of documentation' and 'ease of use' of each assembly tool. These might also be considered important metrics to many end users of such software. We did not assess these qualities and prospective users of such software should be reminded that it might not be straightforward to reproduce any of the assemblies described in this study.

**Practical considerations for *de novo* genome assembly**

Based on the findings of Assemblathon 2, we make a few broad suggestions to someone looking to perform a *de novo* assembly of a large eukaryotic genome:

1. Don't trust the results of a single assembly. If possible generate several assemblies (with different assemblers and/or different assembler parameters). Some of the best assemblies entered for Assemblathon 2 were the evaluation assemblies rather than the competition entries.
2. Don't place too much faith in a single metric. It is unlikely that we would have considered SGA to have produced the highest ranked snake assembly if we had only considered a single metric.
3. Potentially choose an assembler that excels in the area you are interested in (e.g. coverage, continuity, or number of error free bases).
4. If you are interested in generating a genome assembly for the purpose of genic analysis (e.g. training a gene finder, studying codon usage bias, looking for intron-specific motifs), then it may not be necessary to be concerned by low N50/NG50 values or by a small assembly size.
5. Assess the levels of heterozygosity in your target genome before you assemble (or sequence) it and set your expectations accordingly.

# Methods

### Assembly file format

Each assembly was submitted as a single file of FASTA-formatted scaffold sequences which were allowed to contain Ns or other nucleotide ambiguity characters. Submissions were renamed for anonymity and checked for minor errors (e.g. duplicate FASTA headers). Participants were asked to use runs of 25 or more N characters to denote contig boundaries without scaffolds.

### Basic assembly statistics

Basic statistical descriptions of each assembly were generated using a Perl script (assemblathon_stats.pl [33]). The statistics calculated by this script were generated for scaffold and contig sequences (contigs resulted from splitting scaffolds on runs of 25 or more Ns).

### Calculating average vertebrate gene length

Using the Ensembl 68 Genes dataset[70], we extracted the latest protein-coding annotations for human (*Homo sapiens*), chicken (*Gallus gallus*), zebrafish (*Danio rerio*), a frog (*Xenopus laevis*), and a lizard (*Anolis carolinensis*). From these datasets, we calculated the size of an average vertebrate gene to be 25 Kbp.

### CEGMA analysis

The CEGMA tool [38, 40], was used to assess the gene complement of each assembly. Version 2.3 of CEGMA was run, using the --vrt option to allow for longer (vertebrate-sized) introns to be detected.

CEGMA produces additional output for a subset of the 248 most highly conserved, and least paralogous CEGs. For these CEGs, additional information is given as to whether they are present as a full-length gene or only partially. CEGMA scores predicted proteins by aligning them to a HMMER profile built for each core gene family. The fraction of the alignment of a predicted protein to the HMMER profile can range from 20–100%. If this fraction exceeds 70% the protein is classed as a full-length CEG, otherwise it is classified as partial. In both cases, the predicted protein must also exceed a predetermined cut-off score (see[40]).

### Fosmid data

In order to provide an independent reference for assemblies, panels of pooled Fosmid clone Illumina paired-end libraries (~35 Kbp inserts) were made from the bird and snake samples using methods described in [71]. In each case, 10 pools were sequenced at various different pooling levels, from mostly non-overlapping sets of Fosmids (1, 1, 1, 1, 2, 4, 8, 16, 32, and 48 — 96 or 114 clones sequenced in total), generating Illumina 100 x 100 bp paired end sequences, with a predicted insert size of 350 ± 50 bp (observed insert size of 275 ± 50 bp). After adapter and quality trimming using Scythe and Sickle [72], Fosmid reads were aligned using BWA (ver. 0.5.9rc1-2;[73]) to the cloning vector for removal of vector-contaminated read pairs. The Velvet assembler (ver. 1.1.06; [7]) was then used to assemble pools up to 16, at k-mer lengths ranging from ~55 to ~79. Coverage cutoff and expected coverage parameters were set manually after inspecting k-mer coverage distributions, as described in the Velvet manual. Assemblies from higher order pools (those containing reads from more than 16 clones) were highly fragmented, and thus not used in the current work.

**Fosmid analysis**

Repeats in Fosmid sequences were identified using version open-3.3.0 (RMLib 20110920) of the online RepeatMasker software [74]. Reads were aligned to Fosmids using BLASTN with parameters tuned for shorter alignments with some errors (WU-BLASTN [04-May-2006] W=11 M=1 N=-1 Q=2 R=2 kap S=50 S2=50 gapS2=50 filter=dust B=1000000 V=1000000). For snake, 8 lanes of short-read Illumina sequence were used (flowcell ID: 110210_EAS56_0249_FC62W0CAAXX). For bird, we used the 5 lanes of short-insert Illumina data from Duke University (see Supplementary Material for details). Finally, Fosmid sequences were aligned to assembly scaffold sequences using BLASTN with parameters tuned for long, nearly identical alignments (WU-BLASTN [04-May-2006] W=13 M=1 N=-3 Q=3 R=3 kap S=1000 S2=1000).

**VFR COMPASS analysis**

COMPASS [41] is a Perl script that uses Lastz [58] to align assembly scaffolds to a reference, after which the alignment is parsed (using SAMTools [75]) to calculate alignment and coverage island lengths (see Figure 7), which are used to create cumulative length plots (e.g. Figure 10), as well as to calculate coverage, validity, multiplicity, and parsimony metrics. COMPASS was run with a minimum contig length of 200 bp for all submitted assemblies, and with the following lastz command:

```
lastz reference[multiple] assembly[multiple] --ambiguous=n --ambiguous=iupac \
--notransition --step=20 --match=1,5 --chain --identity=98 --format=sam >
out.sam
```

Note the options specifying a minimum 98% identity cutoff for alignments and treatment of 'N' characters as ambiguous (receiving scores of zero, rather than a penalty for mismatch); these and other options may not be appropriate for all cases.

## VFR distance analysis

A Perl script (vfr_blast.pl [33]) was used to loop over successive 1,000 nt regions from the VFR sequences for bird and snake. Pairs of 100 nt 'tag' sequences were then extracted from the ends of each of these regions. All pairs of tag sequences were then searched against all scaffolds for that particular species using BLAST[76]. Matches were only retained if at least 95 nt of each tag sequence aligned to the scaffolds. The resulting BLAST output was processed to determine whether both tag sequences from a pair, matched uniquely to a single scaffold, and if so, at how far apart (expected distance between start coordinates of each tag in a pair is 900 nt).

## Optical Maps

Scaffolds from each assembly were aligned to optical maps that had been generated for each species. Only scaffold sequences from the assemblies that were at least 300 Kbp and possessed at least 9 restriction sites were used for alignment to the optical map supercontigs. The total length of all uniquely aligned sequence was recorded and the resulting alignments were classified into three levels:

Level 1: global alignment, don't allow gaps, strict threshold for score
Level 2: global alignment, allow gaps, permissive threshold for score
Level 3: local alignment, permissive threshold.

Coverage at level 1 reflects situations where the scaffold and optical map are concordant. The second level of coverage (level 2, but not level 1) also reflects situations where the scaffold and optical map are concordant, but where gap sizing or minor differences might lead to lower scores or make it necessary to insert a gap in the alignment. Finally, level 3 coverage (which excludes coverage at levels 1 and 2) represent situations where the global alignment fails, but where the local alignment succeeds. These situations are suggestive of potential chimeric assemblies or a bad join in either the sequence scaffold or optical map. Nonetheless, the regions of the optical maps and sequence that do align are concordant.

## REAPR

All reads were mapped using SMALT version 0.6.2 [77]. All assemblies were indexed using a k-mer length of 13 (-k 13) and step length of 2 (-s 2). Reads were mapped repetitively using the option -r 1. Each read within a pair was mapped independently using the -x flag, so that each read is mapped to the position in the assembly with the best alignment score (regardless of where its mate was mapped). This is critical to the REAPR pipeline, since reads in a pair should not be artificially forced to map as a proper pair when a higher scoring alignment exists elsewhere in the assembly. For short- and long-insert size libraries, the options -y 0.9 and -y 0.5 were used to require 90% and 50% of the reads to align perfectly. The only parameter that was varied when mapping was the -i option to specify the maximum insert size. All BAM files had duplicates marked using the MarkDuplicates function of Picard [78] version 1.67, so that such reads could be ignored by REAPR.

All reads from the two short insert Illumina GAII runs were used for the snake assemblies, with -i 1500. All reads from the 10 Kbp insert library were mapped, using -i 15000. For the fish assemblies, all reads from the fragment and 11 Kbp insert size libraries were mapped using -i 600 and -i 15000 respectively. All reads from the bird BGI short insert Illumina libraries were mapped using -i 1500. Finally, the 20 Kbp insert size Illumina reads were mapped to the bird assemblies with -i 50000.

REAPR version 1.0.12 was used to analyse the assemblies. Perfect and uniquely mapping read coverage was generated by REAPR's *perfectfrombam* function, for input into the REAPR analysis pipeline. This filters the BAM file by only including reads mapped in a proper pair, within the specified insert size range, with at least the given minimum Smith-Waterman alignment score and mapping quality score. Filtering by alignment score ensures that only reads with perfect alignments to the genome were included. The minimum alignment score was chosen to be the read length, since SMALT scores 1 for a match. SMALT assigns a mapping quality score of 3 or below to reads that map repetitively, therefore a minimum score of 4 was used to filter out repetitive reads. The parameters used when running the function *perfectfrombam* for snake, fish and bird were 200 500 3 4 121, 50 250 3 4 101 and 100 900 3 4 150 respectively. Finally, the REAPR pipeline was run on each assembly using the default settings. Due to a lack of coverage of large insert size proper read pairs, it was not possible to run REAPR on the MLK and ABL assemblies in bird and the CTD, CTD*, and CTD** assemblies in fish.

To generate the final summary score for each assembly, within each species the count of error free bases, N50 and broken N50 were normalised as follows. For each of the statistics, the assembly with largest value was given 1 and the remaining values were reported as a fraction of that largest value. For example, if the highest number of error free bases for a particular species was 1,000,000, then all values of error-free bases for that species were be divided by 1,000,000 (so that the best assembly would get a score of 1 for this metric). The same method was applied to the N50 before and after breaking before applying the following formula to calculate a summary score for each assembly:

REAPR Summary Score = Number of error free bases * (broken N50)$^2$ / (original N50)

# Submission of supporting data to the GigaScience database, GigaDB

Three supporting datasets are available for inclusion in GigaDB:

**Dataset 1**

*1) Title:* Assemblathon 2 assemblies

*2) Abstract*

Assemblathon 2 is a genome assembly contest where participating teams attempted to assemble genomes for three vertebrate species using a mixture of next-generation sequencing data. In total, 43 assemblies were submitted for three species (15 for bird, 16 for fish, and 12 for snake). These assemblies were assessed using a wide variety of statistical approaches as well as using experimental data from Fosmid sequences and optical maps.

*3) Author list:* as in publication.

*4) Data types:* Genome assemblies (gzipped FASTA files)

*5) Organism or tissue for each data type:* Budgerigar (*Melopsittacus undulatus*), a Lake Malawi cichlid (*Maylandia zebra*, also referred to as *Metriaclima zebra*), and boa constrictor (*Boa constrictor constrictor*).

*6) Dataset size:* 27 GB (86 compressed *.fa.gz files, and one MD5 checksum file)

*7) README file*

Assemblies were submitted as files of scaffolded contigs with runs of at least 25 consecutive N characters being used to denote gaps between contigs. Scaffolds were split on these runs of N characters to produce contig files. A small number of teams (ABL, CSHL, CTD, and PRICE) provided assemblies which had not undergone scaffolding in which case the contig and scaffold files for each entry are identical. FASTA headers are as originally submitted to the Assemblathon 2 contest. Assembly identifiers (for file names) are described in Table S1 of Additional file 2 from the submitted Assemblathon 2 paper.

*8) Link to data if public, or any related accession numbers in other repositories*

No links required to accession numbers in other repositories.

**Dataset 2**

*1) Title*: CEGMA gene predictions for Assemblathon 2 entries

*2) Abstract*

Assemblathon 2 genome assemblies were assessed for their genic content. This was done by using published tool (CEGMA) that looks for the presence of nearly full-length genes within a single scaffold sequence. Such genes must match HMMs made from a set of 458 highly-conserved genes that are presumed to be conserved in all eukaryotes.

*3) Author list:* Bradnam, K.R., Fass, J.N., and Korf, I.K.

4) *Data types:* FASTA format files for DNA and protein sequences of gene predictions, text files for ancillary information.

*5) Organism or tissue for each data type:* Budgerigar (Melopsittacus undulatus), a Lake Malawi cichlid (Maylandia zebra, also referred to as Metriaclima zebra), and boa constrictor (Boa constrictor constrictor).

*6) Dataset size:* 83.1 MB (compressed files)

*8) README file*

Results were achieved from running CEGMA (v2.4) on all Assemblathon 2 entries.
Only difference from default version of CEGMA was to use the --vrt command-line option to allow vertebrate-sized introns.

One directory exists for each assembly submitted to the Assemblathon 2 contest. Assembly identifiers (for directory names) are described in Table S1 of Additional file 2 from the submitted Assemblathon 2 paper. Each directory contains up to 7 files as follows:

output.cegma.dna - genomic region surrounding each gene prediction
output.cegma.errors - any errors when running CEGMA
output.cegma.fa - protein sequences of the Core Eukaryotic Genes (CEGs), up to 458
output.cegma.gff - exon details in GFF
output.cegma.id - KOG IDs for each CEG present
output.cegma.local.gff -
output.completeness_report - analysis of the 248 most conserved & least paralogous CEGs

*8) Link to data if public, or any related accession numbers in other repositories*

No links required to accession numbers in other repositories.

**Dataset 3**

*1) Title*: Assembled Fosmid sequences used for assessment of Assemblathon 2 entries

*2) Abstract*

Assemblathon 2 genome assemblies for bird and snake were assessed using high-confidence regions of assembled Fosmid sequences. These validated Fosmid regions (VFRs) were included as an additional file as part of the Assemblathon 2 manuscript. This file contains the complete assembled Fosmid sequences for both species (47 sequences for bird, 29 for snake).

*3) Author list:* Fass, J.N., Korf, I.K., Bradnam, K.R., Jarvis, E.D., Shendure, J., Hiatt, J., and Kitzman, J.O.

*4) Data types:* FASTA format file for DNA sequences of assembled Fosmids.

*5) Organism or tissue for each data type:* Budgerigar (Melopsittacus undulatus) and boa constrictor (Boa constrictor constrictor).

*6) Dataset size:* 1.4MB

*8) README file*

Panels of pooled Fosmid clone Illumina paired-end libraries (~35 Kbp inserts) were made from the bird and snake samples using methods described in Kitzman *et al.* 2010 (Haplotype-resolved genome sequencing of a Gujarati Indian individual. *Nature Biotechnology* 2010**, 29**:59). In each case, 10 pools were sequenced at various different pooling levels, from mostly non-overlapping sets of Fosmids (1, 1, 1, 1, 2, 4, 8, 16, 32, and 48 — 96 or 114 clones sequenced in total), generating Illumina 100 x 100 bp paired end sequences. After adapter and quality trimming using Scythe and Sickle, Fosmid reads were aligned using BWA (ver. 0.5.9rc1-2) to the cloning vector for removal of vector-contaminated read pairs. The Velvet assembler (ver. 1.1.06) was then used to assemble pools up to 16, at k-mer lengths ranging from ~55 to ~79.

*8) Link to data if public, or any related accession numbers in other repositories*

No links required to accession numbers in other repositories.

List of Abbreviations

NGS – next generation sequencing
VFR – validated Fosmid region
CLP – cumulative length plot

## Competing Interests

None declared.

# Author contributions

Surnames are provided because initials alone were not specific enough given the large number of co-authors.

### Participating teams

The following teams were responsible for submitting one or more assemblies to the Assemblathon 2 contest. Teams were responsible for downloading input read data, and using various software packages to generate a genome assembly from input data (optionally involving various data pre-processing and quality control steps).

**CSHL team:** P Baranay, S Emrich, MC Schatz; **MLK team:** MD MacManes; **ABL team:** H Chitsaz; **Symbiose team:** R Chikhi, D Lavenier, G Chapuis, D Naquin, N Maillet; **Ray team:** S Boisvert, J Corbeil, F Laviolette, E Godzaridis; **IOBUGA team:** TI Shaw, W Chou; **GAM team:** S Scalabrin, R Vicedomini, F Vezzi, C Del Fabbro; **Meraculous team:** JA Chapman, IY Ho, DS Rokhsar; **Allpaths team:** S Gnerre, G Hall, DB Jaffe, I MacCallum, D Przybylski, FJ Ribeiro, T Sharpe, S Yin; **CBCB team:** S Koren, AM Phillippy; **PRICE team:** JG Ruby; **SOAPdenovo team:** R Luo, B Liu, Z Li, Y Shi, J Yuan, H Zhang, S Yiu, T Lam, Y Li, J Wang; **Curtain team**: M Haimel, PJ Kersey; **CoBiG2 team**; Bruno Miguel Vieira, Francisco Pina-Martins, Octávio S. Paulo; **BCM-HGSC team:** Y Liu, X Song, X Qin, H Jiang, J Qu, S Richards, KC Worley, RA Gibbs; **AbySS team:** I Birol, TR Docking, SD Jackman; **Phusion team:** Z Ning; **CRACS team:** NA Fonseca; **SGA team:** JT Simpson, R Durbin; **Computer Technologies Department (CTD) team:** A. Alexandrov, P. Fedotov, S. Melnikov, S. Kazakov, A. Sergushichev, F. Tsarev **Newber-454 team:** JR Knight.

### Other contributions

S Zhou, S, Goldstein, M Place, DC Schwartz, and M Bechner generated optical maps for each species, performed analysis of each assembly against the optical maps, and contributed to the corresponding sections of the manuscript (including figures).

J Shendure, J Kitzman, and J Hiatt provided Fosmid sequence data for bird and snake.

M Hunt and T Otto performed an analysis of each assembly using the REAPR tool and contributed to the corresponding sections of the manuscript (including figures).

D Earl and B Paten provided analysis of correlation between all key metrics, contributed to the corresponding sections of the manuscript (including figures), and provided additional valuable feedback on the manuscript.

D Haussler helped organize the contest, liaised with various groups to facilitate participation, and assisted with the analysis of correlation between all key metrics.

J Howard, G Ganapathy, and G Zhang provided bird (budgerigar) sequence data.

D Jaffe, K Worley, J Chapman, S Goldstein, M Schatz, D Rokhsar, S Scalabrin, I MacCallum, M MacManes, and S Boisvert provided valuable feedback on the manuscript.

E Jarvis provided bird (budgerigar) sequence data, negotiated with Illumina, BGI, Pacific Biosciences, and Roche 454 to provide additional sequence data, helped organize the contest, liaised with various groups to facilitate participation, and provided valuable feedback on the manuscript.

I Korf helped organize the contest and liaised with various groups to facilitate participation, helped design and coordinate experiments, processed Fosmid sequences to generate Validated Fosmid Regions (VFRs), assessed Fosmids for repeat content and coverage, and provided valuable feedback on the manuscript.

J Fass helped design and coordinate experiments, assembled Fosmid sequences, performed COMPASS analysis of all assemblies using VFR data, and assisted in writing of the manuscript.

K Bradnam helped organize the contest and liaised with various groups to facilitate participation; coordinated submission of entries; made anonymous versions of assemblies available to all third-party groups performing evaluations; helped design and coordinate experiments; calculated and collated all basic statistics for each assembly; ran CEGMA against all assemblies; collected datasets to calculate 'average size of vertebrate gene'; looked for available transcript data; performed distance analysis on VFR sequences; integrated all analyses from other groups; herded goats; performed z-score analyses of assemblies; made heat map from final rankings; drafted, wrote, and edited the manuscript.

# Acknowledgements

The Assemblathon 2 organizers would like to thank all groups and people involved in generating the datasets used for this project. Specifically we thank:

- BGI, Roche 454, Pacific Biosciences, and Erich Jarvis (NIH pioneer award and HHMI funds) for providing the budgerigar (*Melopsittacus undulatus*) sequence data.
- NHGRI for providing funds to sequence Fosmids and produce optical maps.
- The Broad Institute Genomics Platform and Genome Sequencing and Analysis Program, Federica Di Palma, and Kerstin Lindblad-Toh for making the sequence data for the Lake Malawi cichlid (*Metriaclima zebra*) available; and Thomas Kocher at the University of Maryland for providing blood samples for the cichlid.
- Illumina (and Ole Schulz-Trieglaff in particular) for providing the red-tailed boa constrictor (*Boa constrictor constrictor*) sequence data, Matthew M. Hims and Niall A. Gormley for preparing and QC-ing the Illumina mate-pair libraries, and Freeland Dunker at the California Academy of Sciences for taking the snake blood samples.

We would also like to thank the assistance of Amazon Web Services (AWS) and Pittsburgh Supercomputing Center (PSC) for help with the initial distribution of Assemblathon 2 datasets. We also thank the help and support of the UC Davis Bioinformatics Core in providing resources to help finish the analysis of the data.

**Acknowledgements from optical mapping group:**
David Schwartz and the optical mapping team at University of Madison-Wisconsin would like to thank NHGRI for funding for this project.

**Acknowledgements from Fosmid sequencing group:**
This work was supported by National Institutes of Health / National Human Genome Research Institute award R01 HG006283. We thank Alexandra P. Lewis and Ruolan Qiu for assistance with Fosmid library construction.

**Acknowledgements from the REAPR analysis group:**
M. Hunt and T. Otto were supported by the European Union 7th framework EVIMalaR.

**Acknowledgements from Allpaths team:**
This work has been funded in part by the National Human Genome Research Institute, National Institutes of Health, Department of Health and Human Services, under grants R01HG003474 and U54HG003067, and in part by the National Institute of Allergy and Infectious Diseases, National Institutes of Health, Department of Health and Human Services, under Contract No. HHSN272200900018C.

**Acknowledgements from BCM-HGSC team:**
We gratefully acknowledge funding support from a grant U54 HG003273 to RAG from the National Human Genome Research Institute, National Institutes of Health; computational resources provided by National Center For Research Resources, National Institute of Health through a grant to Jeffrey Gordon Reid (1 S10 RR026605) for a massive RAM computer cluster at BCM-HGSC; access to the IBM POWER 7 bioscience computing core BlueBioU machine at Rice University; and an IBM Faculty Award to KCW.  We thank Adam English for mapping the PacBio budgerigar data to the preliminary budgerigar assembly.

**Acknowledgements from CRACS team:**
The work has been partially supported by European Community's FP7/2011-2015, EurocanPlatform under grant agreement no 260791, PTDC/EIA-EIA/121686/2010, and funds granted to CRACS & INESC Tec LA through the Programa de Financiamento Plurianual, Fundacao para a Ciencia e Tecnologia and Programa POSI.

**Acknowledgements from CSHL team:**
National Institutes of Health award (R01-HG006677-12), and National Science Foundation award (IIS-0844494) to MCS

**Acknowledgement for MLK team:**
Matthew MacManes was supported by a National Institutes of Health NRSA Post-Doctoral fellowship (1F32DK093227). Computational resources were supported by the Extreme Science and Engineering Discovery Environment (XSEDE) PSC-Blacklight, which is supported by National Science Foundation grant number OCI-1053575

**Acknowledgements from Meraculous team:**
Work conducted at the US Department of Energy Joint Genome Institute is supported by the Office of Science of the US Department of Energy under contract numbers DE-AC02-05CH11231.

**Acknowledgements from Ray team:**
Computations were performed on the Colosse supercomputer at Université Laval and the Guillimin supercomputer at McGill University (nne-790-ab), under the auspices of Calcul Québec and Compute Canada. The operations of Guillimin and Colosse are funded by the Canada Foundation for Innovation (CFI), the National Science and Engineering Research Council (NSERC), NanoQuébec, and the Fonds Québécois de Recherche sur la Nature et les Technologies (FQRNT). JC is the Canada Research Chair in Medical Genomics. SB is recipient of a Frederick Banting and Charles Best Canada Graduate Scholarship Doctoral Award (200910GSD-226209-172830) from the Canadian Institutes for Health Research (CIHR). FL was supported in part by the Natural Sciences and Engineering Research Council of Canada (grant 262067).

**Acknowledgements from SGA team:**

This work was supported by Wellcome Trust grant WT098051.

**Acknowledgements from SOAPdenovo team:**

We would like to thank the users of SOAPdenovo who tested the program, reported bugs, and proposed improvements to make it more powerful and user-friendly. Thanks to TianHe research and development team of National University of Defense Technology to have tested, optimized and deployed the software on TianHe series supercomputers. The project was supported by the State Key Development Program for Basic Research of China-973 Program (2011CB809203); National High Technology Research and Development Program of China-863 program (2012AA02A201); the National Natural Science Foundation of China (90612019); the Shenzhen Key Laboratory of Trans-omics Biotechnologies (CXB201108250096A); and the Shenzhen Municipal Government of China (JC201005260191A and CXB201108250096A). Tak-Wah Lam was partially supported by RGC General Research Fund 10612042.

# Figure Legends

**Figure 1:** NG graph showing overview of bird assembly scaffold lengths.

The NG scaffold length (see text) is calculated at integer thresholds (1% to 100%) and the scaffold length (in bp) for that particular threshold is shown on the y-axis. The dotted vertical line indicates the NG50 scaffold length: if all scaffold lengths are summed from longest to the shortest, this is the length at which the sum length accounts for 50% of the estimated genome size. Y-axis is plotted on a log scale. Bird estimated genome size = ~1.2 Gbp.

**Figure 2:** NG graph showing overview of fish assembly scaffold lengths.

The NG scaffold length (see text) is calculated at integer thresholds (1% to 100%) and the scaffold length (in bp) for that particular threshold is shown on the y-axis. The dotted vertical line indicates the NG50 scaffold length: if all scaffold lengths are summed from longest to the shortest, this is the length at which the sum length accounts for 50% of the estimated genome size. Y-axis is plotted on a log scale. Fish estimated genome size = ~1.6 Gbp.

**Figure 3:** NG graph showing overview of snake assembly scaffold lengths.

The NG scaffold length (see text) is calculated at integer thresholds (1% to 100%) and the scaffold length (in bp) for that particular threshold is shown on the y-axis. The dotted vertical line indicates the NG50 scaffold length: if all scaffold lengths are summed from longest to the shortest, this is the length at which the sum length accounts for 50% of the estimated genome size. Y-axis is plotted on a log scale. Snake estimated genome size = ~1.0 Gbp.

**Figure 4:** NG50 scaffold length distribution in bird assemblies and the fraction of the bird genome represented by gene-sized scaffolds.

Primary Y-axis (red) shows NG50 scaffold length for bird assemblies: the scaffold length that captures 50% of the estimated genome size (~1.2 Gbp). Secondary Y-axis (blue) shows percentage of estimated genome size that is represented by scaffolds >= 25 Kbp (the average length of a vertebrate gene).

**Figure 5:** Presence of 458 core eukaryotic genes within assemblies.

Number of core eukaryotic genes (CEGs) detected by CEGMA tool that are at least 70% present in individual scaffolds from each assembly as a percentage of total number of CEGs present across all assemblies for each species. Out of a maximum possible 458 CEGs, we found 442, 455, and 454 CEGs across all assemblies of bird (blue), fish (red), and snake (green)

**Figure 6:** Examples of annotated Fosmid sequences in bird and snake.

Panel A = example bird Fosmid, panel B = example snake Fosmid. 'Coverage' track shows depth of read coverage (green = < 1x, red = > 10x, black = everything else); 'Repeats' track shows low-complexity and simple repeats (green) and all other repeats (gray). Alignments to assemblies are shown in remaining tracks (one assembly per track). Black bars represent unique alignments to a single scaffold, red bars represent regions of the Fosmid which aligned to multiple scaffolds from that assembly. Unique Fosmid sequence identifiers are included above each coverage track.

**Figure 7:** Definitions of the COMPASS metrics: Coverage, Validity, Multiplicity, and Parsimony.

**Figure 8:** COMPASS metrics for bird assemblies.

Coverage, validity, multiplicity, and parsimony calculated as in Figure 7.

**Figure 9:** COMPASS metrics for snake assemblies.

Coverage, validity, multiplicity, and parsimony calculated as in Figure 7.

**Figure 10:** Cumulative length plots of scaffold and alignment lengths for bird assemblies.

Alignment lengths are derived from Lastz alignments of scaffold sequences from each assembly to the bird Fosmid sequences. Series were plotted by starting with the longest scaffold/alignment length and subsequently adding lengths of successively shorter scaffolds/alignments to the cumulative length (plotted on y-axis, with log scale).

**Figure 11:** Cumulative length plots of scaffold and alignment lengths for snake assemblies.

Alignment lengths are derived from Lastz alignments of scaffold sequences from each assembly to the snake Fosmid sequences. Series were plotted by starting with the longest scaffold/alignment length and subsequently adding lengths of successively shorter scaffolds/alignments to the cumulative length (plotted on y-axis, with log scale).

**Figure 12:** Short-range scaffold accuracy assessment via Validated Fosmid Regions

First, validated Fosmid regions (VFRs) were identified (86 in bird and 56 in snake, see text). Then VFRs were divided into non-overlapping 1,000 nt fragments and pairs of 100 nt 'tags' were extracted from ends of each fragment and searched (using BLAST) against all scaffolds from each assembly. A summary score for each assembly was calculated as the product of a) the number of pairs of tags that both matched the same scaffold in an assembly (at any distance apart) and b) the percentage of only the uniquely matching tag pairs that matched at the expected distance (± 2 nt.). Theoretical maximum scores, which assume that all tag-pairs would map uniquely to a single scaffold, are indicated by red dashed line (988 for bird and 350 for snake).

**Figure 13:** Optical map results for bird assemblies.

Total height of each bar represents total length of scaffolds that were suitable for optical map analysis. Dark blue portions represent 'level 1 alignments', sequences that were globally aligned in a restrictive manner. Light blue portions represent 'level 2 alignments', sequences that were globally aligned in a permissive manner. Orange portions represent 'level 3 alignments', sequences that were locally aligned. Assemblies are ranked in order of the total length of aligned sequence.

**Figure 14:** Optical map results for fish assemblies.

Total height of each bar represents total length of scaffolds that were suitable for optical map analysis. Dark blue portions represent 'level 1 alignments', sequences that were globally aligned in a restrictive manner. Light blue portions represent 'level 2 alignments', sequences that were globally aligned in a permissive manner. Orange portions represent 'level 3 alignments', sequences that were locally aligned. Assemblies are ranked in order of the total length of aligned sequence.

**Figure 15:** Optical map results for snake assemblies.

Total height of each bar represents total length of scaffolds that were suitable for optical map analysis. Dark blue portions represent 'level 1 alignments', sequences that were globally aligned in a restrictive manner. Light blue portions represent 'level 2 alignments', sequences that were globally aligned in a permissive manner. Orange portions represent 'level 3 alignments', sequences that were locally aligned. Assemblies are ranked in order of the total length of aligned sequence. Note: the SOAP assembly is sub-optimal due to use of mistakenly labeled 4 Kbp and 10 Kbp libraries (see Discussion).

**Figure 16:** REAPR summary scores for all assemblies.

This score is calculated as the product of a) the number of error free bases and b) the squared scaffold N50 length after breaking assemblies at scaffolding errors divided by the original scaffold N50 length. Data shown for assemblies of bird (blue), fish (red), and snake (green). Results for bird assemblies MLK and ABL and fish assembly CTD are not shown as it was not possible to run REAPR on these assemblies (see Methods). REAPR summary score is plotted on a log axis.

**Figure 17:** Cumulative z-score rankings based on key metrics for all bird assemblies.

Standard deviation and mean were calculated for ten chosen metrics, and each assembly was assessed in terms of how many standard deviations they were from the mean. These z-scores were then summed over the different metrics. Positive and negative error bars reflect the best and worst z-score that could be achieved if any one key metric was omitted from the analysis. Assemblies in red represent evaluation entries.

**Figure 18:** Cumulative z-score rankings based on key metrics for all fish assemblies.

Standard deviation and mean were calculated for seven chosen metrics, and each assembly was assessed in terms of how many standard deviations they were from the mean. These z-scores were then summed over the different metrics. Positive and negative error bars reflect the best and worst z-score that could be achieved if any one key metric was omitted from the analysis. Assemblies in red represent evaluation entries.

**Figure 19:** Cumulative z-score rankings based on key metrics for all snake assemblies.

Standard deviation and mean were calculated for ten chosen metrics, and each assembly was assessed in terms of how many standard deviations they were from the mean. These z-scores were then summed over the different metrics. Positive and negative error bars reflect the best and worst z-score that could be achieved if any one key metric was omitted from the analysis. Note: the SOAP assembly is sub-optimal due to use of mistakenly labeled 4 Kbp and 10 Kbp libraries (see Discussion).

**Figure 20:** Correlation between scaffold N50 length and final z-score ranking.

Lines of best fit are added for each series. P-values for correlation coefficients: bird, $P = 0.016$; fish, $P = 0.007$; snake, $P = 0.005$.

**Figure 21:** Parallel coordinate mosaic plot showing performance of all assemblies in each key metric.

Performance of bird, fish, and snake assemblies (panels A–C) as assessed across ten key metrics (vertical lines). Scales are indicated by values at the top and bottom of each axis. Each assembly is a colored, labeled line. Dashed lines indicate teams that submitted assemblies for a single species whereas solid lines indicate teams that submitted assemblies for multiple species. Key metrics are CEGMA (number of 458 core eukaryotic genes present); COVERAGE and VALIDITY (of validated Fosmid regions, calculated using COMPASS); OPTICAL MAP 1 and OPTICAL MAP 1-3 (coverage of optical maps at level 1 or at all levels); VFRT SCORE (summary score of validated Fosmid region tag analysis), GENE-SIZED (the amount of an assembly's scaffolds that are 25 Kbp or longer); SCAFFOLD NG50 and CONTIG NG50 (the lengths of the scaffold or contig that takes the sum length of all scaffolds/contigs past 50% of the estimated genome size); REAPR SCORE (summary score of scaffolds from REAPR tool).

# Tables

**Table 1:** Assemblathon 2 participating team details.

Team identifiers are used to refer to assemblies in figures (Supplementary Table 1 lists alternative identifiers used during the evaluation phase). Sequence data types for bird assemblies are: Roche 454 (4), Illumina (I), and Pacific Biosciences (P). Additional details of assembly software, including version numbers and CPU/RAM requirements of software are provided in Supplementary Table 2. Detailed assembly instructions are available for some assemblies in the Supplementary Methods.

| Team name | Team identifier | Number of assemblies submitted | | | Sequence data used for bird assembly | Institutional affiliations | Principal assembly software used |
|---|---|---|---|---|---|---|---|
| | | Bird | Fish | Snake | | | |
| ABL | ABL | 1 | 0 | 0 | 4 + I | Wayne State University | HyDA |
| ABySS | ABYSS | 0 | 1 | 1 | | Genome Sciences Centre, British Columbia Cancer Agency | ABySS and Anchor |
| Allpaths | ALLP | 1 | 1 | 0 | I | Broad Institute | ALLPATHS-LG |
| BCM-HGSC | BCM | 2 | 1 | 1 | 4 + I + P[1] | Baylor College of Medicine Human Genome Sequencing Center | SeqPrep, KmerFreq, Quake, BWA, Newbler, ALLPATHS-LG, Atlas-Link, Atlas-GapFill, Phrap, CrossMatch, Velvet, BLAST, and BLASR |
| CBCB | CBCB | 1 | 0 | 0 | 4 + I + P | University of Maryland, National Biodefense Analysis and Countermeasures Center | Celera assembler and PacBio Corrected Reads (PBcR) |
| CoBiG[2] | COBIG | 1 | 0 | 0 | 4 | University of Lisbon | 4Pipe4 pipeline, Seqclean, Mira, Bambus2 |
| CRACS | CRACS | 0 | 0 | 1 | | Institute for Systems and Computer Engineering of Porto TEC, European Bioinformatics Institute | ABySS, SSPACE, Bowtie, and FASTX |
| CSHL | CSHL | 0 | 3 | 0 | | Cold Spring Harbor Laboratory, Yale University, University of Notre Dame | Metassembler, ALLPATHS, SOAPdenovo |
| CTD | CTD | 0 | 3 | 0 | | National Research University of Information Technologies, | Unspecified |

|  |  |  |  |  |  | Mechanics, and Optics |  |
|---|---|---|---|---|---|---|---|
| Curtain | CURT | 0 | 0 | 1 |  | European Bioinformatics Institute | SOAPdenovo, fastx_toolkit, bwa, samtools, velvet, and curtain |
| GAM | GAM | 0 | 0 | 1 |  | Institute of Applied Genomics, University of Udine, KTH Royal Institute of Technology | GAM, CLC and ABySS |
| IOBUGA | IOB | 0 | 2 | 0 |  | University of Georgia, Institute of Aging Research | ALLPATHS-LG and SOAPdenovo |
| MLK Group | MLK | 1 | 0 | 0 | I | UC Berkeley | ABySS |
| Meraculous | MERAC | 1 | 1 | 1 | I | DOE Joint Genome Institute, UC Berkeley | meraculous |
| Newbler-454 | NEWB | 1 | 0 | 0 | 4 | 454 Life Sciences | Newbler |
| Phusion | PHUS | 1 | 0 | 1 | I | Wellcome Trust Sanger Institute | Phusion2, SOAPdenovo, SSPACE |
| PRICE | PRICE | 0 | 0 | 1 |  | UC San Francisco | PRICE |
| Ray | RAY | 1 | 1 | 1 | I | CHUQ Research Center, Laval University | Ray |
| SGA | SGA | 1 | 1 | 1 | I | Wellcome Trust Sanger Institute | SGA |
| SOAPdenovo | SOAP | 3 | 1 | 1 | I[2] | BGI-Shenzhen, HKU-BGI | SOAPdenovo |
| Symbiose | SYMB | 0 | 1 | 1 |  | ENS Cachan/IRISA, INRIA, CNRS/Symbiose | Monument, SSPACE, SuperScaffolder, and GapCloser |

[1] BCM-HGSC team also included an evaluation bird assembly with Illumina and Roche 454 data only.

[2] One of the two evaluation assemblies by the SOAPdenovo team included a bird assembly that used Roche 454 and Illumina.

**Table 2.** Overview of sequencing data provided for Assemblathon 2 participants

See Supplementary Methods for a full description of all sequence data.

| Species | Estimated genome size | Illumina | Roche 454 | Pacific Biosciences |
|---|---|---|---|---|
| Bird (*Melopsittacus undulatus*) | 1.2 Gbp | 285x coverage from 14 libraries (mate pair and paired-end) | 16x coverage from 3 library types (single end and paired-end) | 10x coverage from 2 libraries |
| Fish (*Maylandia zebra*)[*] | 1.0 Gbp | 192x coverage from 8 libraries (mate pair and paired-end) | NA | NA |
| Snake (*Boa constrictor constrictor*) | 1.6 Gbp | 125x coverage from 4 libraries (mate pair and paired-end) | NA | NA |

[*] Also described as *Metriaclima zebra* and *Pseudotropheus zebra*